# Evaluating Public Supports to the Investment Activities of Business Firms: A Multilevel Meta-Regression Analysis of Italian Studies

Chiara Bocci[1,2], Annalisa Caloffi[3], Marco Mariani[4] and Alessandro Sterlacchini[5]

[1] *Department of Statistics, Computer Science, Applications "G. Parenti", University of Florence, Italy*
[2] *Florence Center for Data Science, Florence, Italy*
[3] *Department of Economics and Management, University of Florence, Italy*
[4] *IRPET – Regional Institute for Economic Planning of Tuscany, Florence, Italy*
[5] *Department of Economic and Social Sciences, Marche Polytechnic University, Ancona, Italy*

**Abstract**

We conduct an extensive meta-regression analysis of counterfactual programme evaluations from Italy, considering both published and grey literature on enterprise and innovation policies. We specify a multilevel model for the probability of finding positive effect estimates, also assessing correlation possibly induced by co-authorship networks. We find that the probability of positive effects is considerable, especially for weaker firms and outcomes that are directly targeted by public programmes. However, these policies are less likely to trigger change in the long run.

# 1. Introduction

In recent times, awareness has increased about the importance of policies addressing evolutionary or systemic failures and grand societal challenges (Malerba, 2009; Edquist, 2011; Mazzuccato, 2013, 2018). These policies pay particular attention to the innovation system, within which the various skills that are necessary to trigger technological change, learning and innovation can be combined. The system dimension is also necessary to formulate strategies towards societal grand challenges, which require the combination of various types of public and private investments, skills and other resources (Mazzuccato, 2018).

This notwithstanding, the bulk of industrial and innovation policies still consists of the widespread support of individual firms, with their investment projects being neither particularly rooted on a systemic logic, nor focused on grand challenges. Obviously, also the latter type of policies has its own *raison d'être*, rooted in the ideas of market and managerial failures, which might prevent firms, especially SMEs, from investing in innovation (e.g., Metcalfe, 1995; Peneder, 2008). However, their effectiveness is the subject of constant debate in several corners of the world (Rodrik, 2008; Cimoli et al., 2009; Pianta et al., 2020).

Fortunately, the culture and practice of counterfactual programme evaluation has been on the rise in recent years, which makes it more and more possible to formulate judgments based on evidence, rather than ideology. The main results of these assessments can be summarised using a meta-regression analysis (MRA), which is "the systematic review and quantitative synthesis of empirical economic evidence on a given hypothesis, phenomenon, or effect" (Stanley et al., 2013, p. 391). The potential usefulness of this approach is particularly marked in the area of programme evaluation, where causal programme effects often suffer from limited external validity (Olsen et al., 2013; Alcott, 2015; Athey and Imbens, 2017). MRA can help generalise beyond "local" inferences (Bandiera et al., 2016; Vivalt, 2020) and understand to which extent, and in which situations, enterprise and innovation policies are effective according to the estimates reported in counterfactual evaluations.

Despite the importance and diffusion of enterprise and innovation policies, few MRAs exist in the literature which focus on specific policy tools, such as R&D subsidies, R&D tax-credits or both (Garcia-Quevedo, 2004; Castellacci and Mee Lie, 2015; Dimos and Pugh, 2016). Our paper contributes to this literature in multiple, original ways.

First, we do not look only at R&D subsidies and R&D tax-credits. We broaden the scope of our MRA to cover counterfactual estimates related to wider range of public supports to the investment process of private firms: subsidies, tax-credits and direct loans for R&D, as well as subsidies, direct loans and public loan guarantee schemes in favour of more generic investments. Moreover, unlike previous studies, we strongly focus on features such as the type of beneficiaries of the policies (e.g. large vs small firms), the level of government at which the incentives are granted, or the time at which an effect is likely to be found.

Second, to tackle issues of publication bias, we perform a systematic search of the available evidence and we collect both published and grey (i.e. unpublished) literature. Indeed, the validity of the conclusions reached by a MRA can be challenged by the preference of journal editors to publish studies that report conclusive results, which might imply that studies with significant estimates are over-represented in the published literature. This issue can be addressed by including in the MRA as many studies as possible that appeared outside of journals (Hopewell et al., 2007; Card et al., 2010).

In fact, statistically rigorous evaluation analyses can be found also in policy reports that are not intended for publication in international scientific journals, but which, nonetheless, contain valuable information. It is evident that the systematic research of grey literature requires to delimit the field. We choose to focus on a single country, Italy, in order to guarantee that the studies considered all refer to a relatively homogenous institutional and business environment. Despite the strong diffusion of evaluation culture and practice, Italy has been vastly disregarded by the existing meta-analyses. For example, our search reveals that from 2000 to 2016 the Italian counterfactual program evaluation literature on firms' investments support consists of no less than 50 studies, including 1,066 treatment effect estimates, 564 of which are related to R&D and 502 to other investment support programmes.[1]

Third, to fully acknowledge the hierarchical structure of the data, we build on the meta-regression approach by Card et al. (2010, 2017) and specify a multilevel meta-regression model for the probability of having a positive effect that is also statistically significant. Indeed, our data are laid in a hierarchical structure, with treatment effect estimates at the lower level and studies at the upper level. Estimates may depend not only on the characteristics of the programme under investigation and its beneficiaries, but also on the choices made by the authors in carrying out their studies. Both aspects may be partially unobservable, especially when related to intrinsic programme quality or to particular analytical choices that the authors choose not to reveal. Disentangling possible observable sources of success, while accounting for the influence exerted by unobservable factors, can be useful to understand how, and for whom, programmes may be improved, as well as to ease learning by policymakers (Mytelka and Smith, 2002; McKelvey and Saemundsson, 2018). To this end, the usage of hierarchical statistical models for MRA that include terms of unobserved heterogeneity seems particularly appropriate. Nonetheless, this class of models has found very limited application in the area of economic MRA (Awaworyi Churchill et al., 2016; Ugur et al., 2016; Ugur et al., 2017).

Fourth, we acknowledge that the programme evaluation literature that is meta-analysed may be characterised – as in many other scientific fields – by a networked structure that sees some scholars regularly publish on the topic, while others contribute more episodically to the field literature (Newmann, 2001). We believe that such structure, ignored by earlier MRA, deserve to be accounted for. In particular, when using multilevel models, it could pose a threat to the plausibility of the standard assumption of between-group independence, in that two or more articles by the same author might share some unobserved "scholar effect". We argue that this issue should be addressed, at least during robustness analysis, by introducing sensible hypotheses about the correlation structure that might link study-level random components.

Finally, to assess the threat posed by *p*-hacking, or selective reporting – such that authors are more likely to report the estimates that satisfy the minimal requirements of statistical significance – we borrow manipulation tests from regression discontinuity designs (McCrary, 2008; Cattaneo et al., 2017), which may suggest the existence of discontinuity in the density of estimates at the two sides of conventional thresholds of statistical significance.

Section 2 presents how we have collected Italian counterfactual estimates into a single dataset, while Section 3 is devoted to the presentation of our hierarchical meta-regression model. Section 4 outlines how the assumption of independence between studies may be relaxed. Section 5 reports the results of our hierarchical MRA and Section 6 concludes.

---

[1] Note that existing meta-analyses in this field, which have a global coverage, include almost this same number of studies and estimates, with only one or two studies that are related to Italy.

## 2. Data

To collect a relevant sample of counterfactual evaluation studies on the effectiveness of public incentives to the investment activities of Italian firms, we started with a literature search on Google, Google Scholar, EconLit, IDEAS, Scopus and ISI Web of Science, by using the keywords "enterprise policy evaluation", "R&D policy evaluation", "innovation policy evaluation" (also in Italian: "valutazione politiche per le imprese", "valutazione politiche per la R&S", "valutazione politiche per l'innovazione"). Once this initial list was created, we selected only those studies that were related to Italian enterprise or innovation policies, implemented both at national and regional scale.[2] Then, we carefully investigated the reference lists of the studies retrieved and searched for papers that were not already included in our initial selection. To complete our list, we asked information to colleagues affiliated to three major Italian associations of economists, which often host sessions devoted to enterprise policy evaluation in their conferences.[3] We completed our search in December 2016.To facilitate the comparability of the studies, we selected only those papers adopting the methodological tools of the econometrics of programme evaluation (Imbens and Wooldridge, 2009) or other methodologies that are suitable to draw causal claims (e.g. structural models, marginal structural models, etc.). Since these methods were primarily thought for estimating treatment effects in the presence of independent observations (e.g. under the Stable Unit Treatment Value Assumption; Imbens and Rubin, 2015), they have been mostly used to evaluate the incentives to individual enterprises, rather than those targeting consortia of firms or other types of temporary associations. Therefore, we restrict attention to studies on individual firm incentives.

As a result of this search process, we have 50 selected studies, which were published (or written) from 2000 to 2016 (see Web Supplementary Material). Only 18 of these studies are written exclusively in Italian, while the rest is written in English (or in both languages). Studies are both articles published in refereed academic journals, and book chapters or unpublished manuscripts (e.g., working papers or policy reports). The choice of including studies appeared in outlets other than scientific journals was made not only for the sake of completeness, but also to guard against publication bias.[4] Despite our best efforts to cover the whole relevant literature, we must consider the possibility that some existing studies dedicated to the evaluation of programmes implemented in the time period under analysis have been involuntarily overlooked. It is also possible that such studies appear during the writing of this paper or will come out one day. Given this possibility, we must look at the selected studies (and at the treatment effect estimates they report) as if they were a large sample from a super-population of Italian studies (of estimates). Each of the 50 studies includes one or more treatment effect estimates, as well as a description of the policy under analysis. To create the database for our meta-analysis, we carefully read the studies and agreed on how to codify the relevant information. In particular, we adopted the following protocol. We selected a subset of 10 articles, which we all read and discuss how to codify. Then, each co-author codified another 10 articles and her or his work was reviewed by a different co-author. In the final stage, each co-author reviewed the complete database.

---

[2] Italy is characterised by a quasi-federal system in which a large part of enterprise and innovation policies are shared between Regions and the State on the basis of the principle of vertical subsidiarity (Caloffi and Mariani, 2018). As a result, regional-scale initiatives coexist with some programmes of national relevance that are managed by the Italian government.

[3] We interviewed our colleagues during the annual meetings (2016) of the SIE-Italian Economic Association, SIEPI-Italian Society of Industrial Economics and Policy, AISRe-Italian Association of Regional Science.

[4] During the preparation of this paper, some of the working papers initially selected ended up being published in a journal. In such cases, we updated the records relative to the study in our dataset. No update occurred for papers that appeared in journals during 2017.

Often, the studies reported treatment effect estimates on multiple outcome variables. We looked at all estimates, obtained under the classical binary-treatment framework, which the authors chose to include in the section(s) devoted to results, leaving aside only those presented in sections or appendixes dedicated to robustness checks or sensitivity analysis.[5] In some cases, the authors chose to present in their results section more than one treatment effect estimate on the same outcome variable, without stating any order of preference. For example, some papers adopting regression discontinuity designs report estimates under different bandwidths and/or different polynomial approximations, while others that perform statistical matching may report estimates under different numbers of matched controls. When this occurred, we selected the estimate that, based on statistical theory, was less likely to be affected by bias (e.g., the one associated with the narrowest bandwidth or the one associated with the lowest number of matched controls, e.g., Imbens and Wooldridge, 2009; Gelman and Imbens, 2019), whereas estimates on different outcomes were all kept in. The result of our selection process is a hierarchical database including 1,066 estimates from 50 studies. On average, each article reports 21.3 estimates (standard deviation = 31.1). Half of these estimates (10.6 on average) refer to the overall treatment effect, while 10.7 refer to treatment effects for specific subgroups of firms. The estimates are often related to several outcome variables, which may be expressed in different measurements units. As a whole, we have found more than one hundred different outcome variables. Those that are more frequently used are employment (about 8% of estimates), turnover (7%), investments, R&D expenditures or R&D employees, value added, productivity, probability of survival and profitability. These outcomes are sometimes expressed in levels, while other times they are ratios, variations or growth rates. As will be explained later in the paper, this heterogeneity of measurement units call for some transformation of the treatment effect estimates in order to make them comparable.

Table 2 reports descriptive statistics related to the variables characterising the estimates and the studies from which they are drawn. Such variables are primarily classified depending on how often they take values that are constant at the study level. If a variable never changes at the study level but only between studies, then it can be viewed as a variable describing the study. If it also varies within the study, then it is related to the estimate level.

## 3. Methodology

Our data have a two-level hierarchical structure. Let $i$ denote the $i$-th collected treatment effect estimate (first level of the hierarchy, i=1,…,n) drawn from study $j$ (second level, j=1,…,J). Let $y_{ij}$ be the value of such estimate or some reasonable transformation of this value, and $x_{ij}$ the vector of covariates related to such estimate. As shown in Table 2, some of these variables never change across the estimates from the same study, whereas others do. To properly account for the hierarchical structure of the data, we resort to a multilevel meta-regression model; an approach that is still underused in economic meta-analysis studies in spite of its potential (e.g., Awaworyi Churchill et al., 2016; Ugur et al., 2016; Ugur et al., 2017). In very general terms, the response variable in such model is function of both the observed explanatory variables and a term of unobserved heterogeneity at the study level, $u_j$. In brief, this approach entails that the variability that remains unexplained by covariates is captured by two different error components: one associated with the unobserved factors that all estimates grouped in a given study $j$ have in common, and one related to the individual level.

---
[5] An extremely limited number of papers also reported estimates obtained in a continuous-treatment framework, for example using generalised propensity scores and dose-response functions. These latter few estimates were left out of the sample, as – for several reasons – they were hardly comparable to the others.

### 3.1 The outcome variable

As the collected estimates are expressed in many heterogeneous measurement units, we need to transform them so that the response variable of our meta-regression model has one single measurement unit. To do so, we divide the raw value of each estimate by its associated standard error, thus obtaining the *t*-statistic $t_{ij}$. We recode the sign of the $t_{ij}$ in those cases where a negative sign goes in the direction desired by the policy, and vice versa. For example, if public support reduces the risk of firm exit, then the negative sign of the *t*-statistic must be turned positive; instead, if it increases exit risk, then the positive sign of the $t_{ij}$ must be turned negative.[6]

The use of $t_{ij}$ as an outcome variable does not account for the fact that there is enough support against the null hypothesis of null that the average treatment effect is equal to zero only for when $t_{ij}$ is above a certain value. In other words, only *t*-statistics above a certain threshold denote positive effects that are conventionally regarded as being statistically significant. This can be accounted for by creating a discrete response variable for both the sign and the statistical significance of the treatment effect estimate. For example, in their meta-regression of causal studies on active labour market policies, Card et al. (2010; 2017) create an ordinal response variable whose three values denote, respectively, statistically significant negative effects, insignificant effects, and significant positive effects. We will adopt this same approach but, given the negligible number of statistically significant negative effects reported in the pool of studies under investigation (only 5.8%, see Table 2), it seems sensible here to construct a simpler binary response variable that takes the value of one if the estimate is both positive and statistically significant, and zero otherwise (see also Garcia-Quevedo, 2004; Kluve, 2010). A meta-regression model with such a response variable is actually a model for the probability of having a positive $t_{ij}$ greater than the critical value guaranteeing the desired level of statistical significance.

All tests reported in the studies under investigation are two-tailed, i.e. they test the null hypothesis that the effect is zero *vs.* the alternative hypothesis that it differs from zero. Estimates that, according to such two-tailed tests, are significant at a 10% level are usually viewed as worthy of some interest. However, since we focus only in significant positive effects, we consider the right-tailed test for the null hypothesis that the effect is null or harmful to firms *vs.* the alternative hypothesis that it is beneficial, and transform the original two-tailed tests accordingly.

---

[6] Other options to transform the value of the estimates are partial correlation coefficients and elasticities (Stanley and Doucouliagos, 2012). Such options do not seem suitable to our context of analysis, which is characterised by estimates obtained under the classical binary-treatment framework and with a widespread use of semi-parametric methods that try to avoid model dependence.

Thus, the response variable of our meta-regression model is defined as follows:

$$y_{ij} = \begin{cases} 1 & \text{if } t_{ij} > 1.645 \\ 0 & \text{otherwise} \end{cases}$$

where 1.645 is the critical value for the right-tailed test being significant at 5%.

As shown in Table 2, 32.2% of estimates are positive and associated with *p*-values that do not exceed 5%.

For sensitivity analysis purpose, we will also consider a second binary outcome variable for results that are both positive and significant at a 2.5% level according to a right-tailed test ($t_{ij} > 1.96$). These represent 25.4% of all estimates. This allows us to guard against possible practices of *p*-hacking that might occur in the proximity of the 5% significance threshold (e.g., Brodeur et al., 2016; Bruns, 2017).

## 3.2 Assessing the threat posed by *p*-hacking

*P*-hacking denotes the authors' choice to report only statistically significant estimates that confirm the hypotheses of interest and may translate into an inflation of just-rejected tests for the null hypothesis of no average effect, possibly due to unobservable, *ad hoc* statistical practices. To assess the presence of p-hacking, in Figure 1 we show the Kernel probability density function of the *t*-statistics in the region $0 \leq t_{ij} \leq 3$, which includes three major threshold values for statistical significance. A ditch appears just below the 1.645 threshold, followed by a hump, which might raise the suspect of *p*-hacking around the 5% significance cutoff (10% with usual two-tailed tests).

Building on Gerber and Malhotra (2008), we investigate further the presence of p-hacking using a "manipulation test" based on density discontinuity, which is borrowed from the methodological literature on regression discontinuity designs. The idea behind such tests is that, in the absence of any manipulation around the threshold, the density should be continuous at the threshold itself. In particular, we apply the test developed by Cattaneo et al. (2018) based on a local polynomial density estimator[7], which builds on McCrary (2008).

As shown in Table 1, a global test on all available estimates does not reject the null hypothesis that there is no discontinuity in the densities at the two sides of the cutoff, though Figure 1 might suggest the contrary. After we focus on meaningful subgroups of estimates, we find support in favour of a discontinuity at 1.645 only for the subset of estimates that were published, whereas no jumps are found at higher significance thresholds. The jump we find for published estimates does not constitute a proof that *p*-hacking has occurred. However, it suggests that including many estimates that are not drawn from journals in our sample was the right call, and that a sensitivity analysis using a significance threshold of 2.5% is appropriate[8].

---

[7] The left and right approximations of the density at the threshold are done independently from each other. Inference relies on a local cubic (triangular) Kernel approximation, with bandwidths optimised separately at each side using a local quadratic fit.

[8] Since the observed power of a given $t_{ij}$ is a one-to-one function of its own *p*-value, $p_{ij}$ (Hoenig and Heisey, 2001), repeating the meta-analysis with a smaller significance threshold is equivalent to see what happens if one (as in

*Figure 1. Probability density function of the t-statistics in the region* $0 \leq t_{ij} \leq 3$

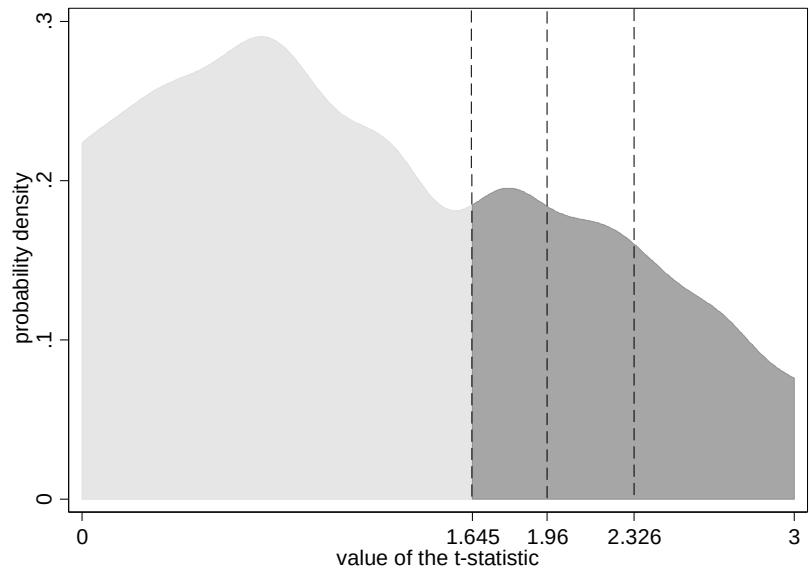

Notes. The area of rejection of the null hypothesis is dark grey. Smoothing was obtained through a Gaussian Kernel with bandwidth = 0.15

*Table 1. Manipulation tests based on density discontinuity at selected threshold values of the t-statistic*

| Estimates come from | 5% significance Threshold = 1.645 | | 2.5% significance Threshold = 1.960 | | 1% significance Threshold = 2.326 | |
| --- | --- | --- | --- | --- | --- | --- |
| | test statistic | *p*-value | test statistic | *p*-value | test statistic | *p*-value |
| All studies | 1.263 | 0.207 | -0.723 | 0.470 | 0.092 | 0.927 |
| Parametric approach | 1.109 | 0.267 | -1.384 | 0.166 | 0.448 | 0.654 |
| Semi-parametric approach | 1.057 | 0.291 | 0.515 | 0.607 | -0.505 | 0.613 |
| Published studies | 2.229** | 0.026 | -0.452 | 0.651 | -0.710 | 0.477 |
| Studies appeared elsewhere | 0.317 | 0.751 | -0.892 | 0.372 | 0.797 | 0.426 |

\* p<0.10; \*\* p<0.05; \*\*\* p<0.01

### 3.3 The meta-regression model

Our multilevel approach builds on Card et al. (2010; 2017) and places their approach in a multilevel framework. In so doing, our work differs from the multilevel MRAs conducted by Awaworyi Churchill et al. (2016), Ugur et al. (2016) and Ugur et al. (2017), who build on the Stanley's approach to meta-regression (Stanley and Doucouliagos, 2012).

---

Ioannidis et al., 2017) is more demanding in terms of the statistical power that each significant estimate should exhibit to deserve consideration. In our study, the positive treatment effect estimates that are significant at 5% have a median observed power of 81.7%, a minimum of 50.3% and a maximum near to 100%. By selecting from the previous estimates only those whose $p_{ij} < 0.025$ we conduct the analysis on a subset of significant estimates that have more power. Here, the median observed power is 87.3% and the minimum is 62.5%.

We specify the following multilevel meta-regression model for the *logit* of the probability of having a significantly positive treatment effect estimate:

$$\eta_{ij} = \text{logit}\left[\Pr(y_{ij}=1 | x_{ij}, u_j)\right] = \beta_0 + \beta_x x_{ij} + u_j, \quad [1]$$

where $x_{ij}$ is a set of $p$ explanatory variables of interest, $\beta_x$ is the vector of related unknown coefficients and $u_j$ is the study random coefficient. Let define $D$ as a $n \times J$ matrix with element $d_{ij}$ taking value 1 if observation $i$ is in study $j$ and 0 otherwise, and let assume vectors $\eta = [\eta_{ij}]$, $\beta = [\beta_0, \beta_x]$, $u = [u_j]$ and matrix $X = [x'_{ij}]$, then model [1] in matrix notation is

$$\eta = X\beta + Du. \quad [2]$$

The set of covariates $X$ should include, in addition to other estimate- and study-level variables, a covariate that measures sample size (e.g., the square root of the number of observations) or the estimates' precision (e.g., the estimates' standard error, which depends on the sample size), in order to evaluate and control for the publication bias that might be due to this source (Stanley and Doucouliagos, 2012).

The term of unobserved heterogeneity $u_j$ could be defined as a fixed parameter or as a random term. Since we look at our studies as at a sample from a population of Italian evaluation studies and wish to draw conclusions pertaining to this population, and since we also wish to estimate coefficients associated to study-level explanatory variables, then it is appropriate to view the term $u_j$ as a random coefficient (Snijders and Bosker, 2012, Chapter 4).

We initially assume that each random coefficient is independent and identically distributed and follows a Normal distribution with zero mean and unknown variance $\sigma_u^2$: $u \sim N(0, \sigma_u^2 I_J)$, where $I_J$ denotes the $J \times J$ identity matrix. The assumption of independence, which is standard within multilevel models, will be relaxed later in the paper. We also hypothesise that the random coefficients are uncorrelated with the estimate-level explanatory variables conditional on the study means (or proportions) of the explanatory variables themselves. Modelling such dependence explicitly guarantees the unbiased estimation of $\beta$ (Mundlak, 1978; Skrondal and Rabe-Hesketh, 2004; Snijders and Bosker, 2012). Model fitting is carried out by maximum likelihood estimation.

Each coefficient in vector $\beta$ represents the change in the log-odds associated with a one-unit change in the corresponding predictor, conditional on the term $u_j$. The direct influence of each covariate on the probability of having a significantly positive treatment effect estimate can then be computed as follows:

$$\Pr(y_{ij}=1 | x_{ij}, u_j) = \frac{\exp(\beta_0 + \beta_x x_{ij} + u_j)}{1 + \exp(\beta_0 + \beta_x x_{ij} + u_j)}$$

If one is interested in probability predictions that are free of the term of unobserved study heterogeneity, these can be obtained by fixing $u_j$ at its expected value of zero.

### 3.4 Model specification

A key decision relates to the specification of the meta-regression model, both in terms of the covariates to be included and of the functional form that the linear predictor should take.

Building on previous literature, we select the following explanatory variables, which are displayed in Table 2. First, to describe the type of programme, we define a dummy variable which takes value 1 for supports aimed at investments and value 0 for R&D or innovation supports; a categorical variable describing the type of incentive provided to firms (direct loan; loan guarantee; non-repayable subsidy; tax credit; support not specified by the authors or mixed); and a categorical variable for the level of government implementing the programme.[9] For the latter variable, in addition to the categories 'national' and 'regional', we also define an 'unknown' category, as some of the studies we analyse are based on survey data that do not report this information.

A second group of variables accounts for study characteristics. Here, a dummy indicates if a study is published in refereed journals or book chapters rather than in working papers or research reports; and a categorical variable describes the methodology adopted for estimation (parametric DID; RDD; matched DID; matching; other parametric methodologies). Sample size, on which publication bias is often believed to depend (Begg and Berlin, 1988), is accounted for by a variable reporting the square root of the observations constituting the largest sample in each study. The use of the square root of sample size is advised by several scholars (e.g., Stanley, 2005; Card et al., 2010) and our choice to consider the largest sample in each study is motivated by the idea that, when studies report combinations of estimates based on both the available full sample and on some subsamples of particular interest, publication selection is likely to depend more on the size of the full sample rather than on that of its possible partitions.

Third, we define three variables describing the outcome considered and the related type of effect that is estimated in the studies. The first is a dummy taking the value of 1 if the outcome variable is a quantity that is directly targeted by the incentive provided by the evaluated programme, and 0 if it refers to an outcome that is more likely to be affected by the incentive in an indirect fashion, only if something else occurs (or does not occur) in the meantime. An example can help clarify this distinction. Let us consider a public loan guarantee. An estimate of the effect of the programme on the reduction of the interest rate on aggregate debt refers to a quantity that is directly targeted by the policy. On the contrary, an estimate related to firms' turnover or productivity growth refers to outputs that may be triggered by the innovation process itself, but which are not the direct target of the programme. The second of such variables is a dummy taking the value of 1 if the outcome variable is measured by the author of the estimate after the firm participates in the programme providing supports, rather than during its participation. This information is important, as some outcome variables can be expected to change very soon after the receipt of a given support, while others could take more time to change. For example, in the case of R&D subsidies, R&D expenditures and firms' propensity to co-operative research can change immediately after receipt of the subsidy, while the effects of the policy (if any) on firms' patenting activity or profitability can reasonably be seen only after some time.[10] The third and last of such variables is a categorical one for the type of firms to which the estimate refers. Often, in addition to estimates referring to all participant firms, studies also report estimates for specific subgroups chosen by the authors. Depending on the type of programme and on the market failure it tries to address, we classified all estimates in four categories: all firms with no distinctions; estimates relative to the subgroup of disadvantaged (or weaker) firms; estimates relative to the subgroup of advantaged firms; estimates relative to other subgroups of firms. Disadvantaged firms are small firms, newborn firms, credit constrained firms, firms with no R&D experience and the like, whose investing activity, according to the literature, is likely to be hindered by certain obstacles (e.g., Berger and Udell, 1998; Beck and Demirguc-Kunt, 2006; Peneder, 2008; Storey et al., 2016). On the contrary, advantaged firms are larger firms, firms that do not have any credit constraints, firms with R&D experience, and so forth. Sometimes, the estimates reported in the studies refer to other subgroups, such as firms that are located in a particular geographical area or firms that operate in a particular sector. Since the definition of these subgroups does not respond to a general logic (as it is with the above mentioned definition), but is rather the reflection of a specific interest of the author(s) in that particular study, we group all these latter estimates in a residual category.

---

[9] In the studies included in our analysis, programmes aimed at R&D may employ the following instruments: subsidies, direct loans and tax-credit. Programmes aimed at investments may employ the following instruments: subsidies, direct loans, tax-credit and public loan guarantees.

[10] Out of 431 estimates on outcomes that are measured simultaneously to programme participation, 39% refer to outcomes that are directly targeted by that particular type of programme, whereas 61% refer to outcomes that might be affected by the programme in a more indirect fashion. Out of 635 estimates on outcomes that are measured after programme participation, 124 (19.5%) refer to outcomes that are directly targeted by that particular type of programme, whereas 80.5% refer to outcomes that might be affected by the programme in a more indirect fashion.

Finally, in order to control any possible systematic differences related to time, we include a dummy variable which indicates if estimates are related to programmes implemented (or survey data collected) before the recent economic crisis rather than during the crisis.

In addition to the previous explanatory variables that may relate either to the estimates or to the study the estimates come from, we use additional explanatory variables with the mean (proportion) of the estimate-level covariates in each study to guarantee independence between random coefficients and estimate-level regressors.

It makes sense to consider these additional descriptors of the study context provided there is non-negligible variability of the underlying estimate-level covariates within the studies themselves. Motivated by the statistics reported in Table 2, we added to the model the study-level proportions of: the dummy for the timing of the effect; the dummy for the type of outcome variable; and the categorical variable for the type of firms to which the estimate refers.

For each discrete variable mentioned so far, Table 3 reports the proportion of significantly positive treatment effect estimates and the average *t*-statistic associated to all estimates that fall under each level of these variables. These are just additional descriptive statistics in that such "vote counts" are not suitable, *per se*, to establish which programme, estimate or programme characteristics are associated with higher probability of success.

With respect to the functional form of the predictor, the main point is to assess whether it is sufficient to insert covariates in the model in a merely additive fashion or if, instead, the inclusion of some interaction terms between covariates ensures a better fit to the available data. Economic reasoning may provide useful guidance in this process, by suggesting interactions that might make sense in our setting, such as those between the aim of the programme, the incentive type, the government level, the type of outcome variable, its timing and the kind of firms the estimate refers to. From a statistical perspective, such an assessment requires to evaluate whether the coefficients associated with interacted covariates are non-negligible and to check if the inclusion of interacted covariates leads to significant gains in the likelihood of the model. After a careful evaluation of interactions in the data at hand, we found that none of these fulfil the two previous criteria. Therefore, we must conclude that the insertion of covariates in an additive fashion is appropriate.

*Table 2. Some descriptive statistics of the studies and estimates considered in the meta-regression analysis*

| | At the level of estimates | | No. of studies in which the variable is constant across estimates | At the level of studies | |
|---|---|---|---|---|---|
| | **Proportion/ Mean** | **S.D.** | | **Proportion/ Mean** | **S.D.** |
| Treatment effect is significantly (5%, left-tailed) negative (1/0) | 0.058 | 0.234 | 33/50 | 0.063 | 0.158 |
| Treatment effect is significantly (5%, right-tailed) positive (1/0) | 0.322 | 0.467 | 13/50 | 0.526 | 0.327 |
| Treatment effect is significantly (2.5%, right-tailed) positive (1/0) | 0.254 | 0.436 | 13/50 | 0.467 | 0.342 |
| *t* statistic (cont.) | 1.117 | 3.394 | 5/50 | 1.795 | 2.352 |
| ***Variables that are always constant within studies*** | | | | | |
| Programme aimed at investments (1/0, base: aimed at R&D) | 0.471 | 0.499 | 50/50 | 0.520 | 0.505 |
| Study was published (1/0, base: unpublished) | 0.588 | 0.492 | 50/50 | 0.740 | 0.443 |
| Sample size (No. of observations, cont.) | 3,522 | 10,475 | 50/50 | 4,467 | 13,336 |
| ***Variables that are usually constant within studies*** | | | | | |
| Programme implemented before the recent crisis (1/0, base: during crisis) | 0.568 | 0.496 | 46/50 | 0.696 | 0.448 |
| *Government level* | | | | | |
| national | 0.368 | 0.482 | 48/50 | 0.420 | 0.488 |
| regional | 0.577 | 0.494 | 49/50 | 0.430 | 0.495 |
| unknown (survey data) | 0.055 | 0.229 | 49/50 | 0.150 | 0.354 |
| *Incentive type* | | | | | |
| direct loan | 0.155 | 0.362 | 46/50 | 0.065 | 0.216 |
| loan guarantee | 0.053 | 0.223 | 50/50 | 0.100 | 0.303 |
| subsidy | 0.644 | 0.479 | 47/50 | 0.610 | 0.479 |
| tax credit | 0.044 | 0.205 | 49/50 | 0.083 | 0.274 |
| unspecified or mixed | 0.105 | 0.307 | 48/50 | 0.142 | 0.340 |
| *Methodology* | | | | | |
| Difference in differences (parametric) | 0.141 | 0.348 | 49/50 | 0.145 | 0.350 |
| Regression discontinuity design | 0.129 | 0.335 | 49/50 | 0.155 | 0.360 |
| Matched difference in differences | 0.299 | 0.458 | 49/50 | 0.261 | 0.442 |
| Matching | 0.303 | 0.460 | 47/50 | 0.282 | 0.442 |
| Other (parametric) | 0.129 | 0.335 | 48/50 | 0.157 | 0.354 |
| ***Variables that are seldom constant within studies*** | | | | | |
| Directly targeted outcome (1/0, base: indirectly targeted outc.) | 0.274 | 0.446 | 32/50 | 0.354 | 0.415 |
| Non simultaneous effect (1/0, base: simultaneous) | 0.596 | 0.491 | 41/50 | 0.460 | 0.462 |
| *Estimate refers to* | | | | | |
| all firms (grand ATE or ATT) | 0.498 | 0.500 | 30/50 | 0.704 | 0.358 |
| disadvantaged firms | 0.172 | 0.377 | 35/50 | 0.104 | 0.200 |
| advantaged firms | 0.114 | 0.319 | 39/50 | 0.048 | 0.103 |
| other subgroup of firms | 0.216 | 0.412 | 35/50 | 0.144 | 0.261 |
| No. of observations | 1,066 | | | 50 | |

*Notes*. Group mean refers to the between-study mean of the within-study means. All variables, with the sole exception of *n. of firms involved in estimation* are binary variables.

*Table 3. Proportions of significantly positive estimates and mean t-statistic for selected variables*

|  | No. of estimates | Proportion of positive estimates for which | | *t*-statistic | |
|---|---|---|---|---|---|
|  |  | $t_{ij} > 1.645$ | $t_{ij} > 1.96$ | Mean | S.D. |
| Programme aimed at R&D | 564 | 0.287 | 0.215 | 0.898 | 1.722 |
| Programme aimed at investments | 502 | 0.361 | 0.299 | 1.364 | 4.587 |
| Study was published | 627 | 0.311 | 0.260 | 1.172 | 4.160 |
| Study appeared in other outlet | 439 | 0.337 | 0.250 | 1.040 | 1.807 |
| Programme implemented before the recent crisis | 606 | 0.285 | 0.226 | 1.209 | 4.289 |
| Programme implemented during the recent crisis | 460 | 0.37 | 0.291 | 0.997 | 1.569 |
| *Government level* | | | | | |
| national | 392 | 0.293 | 0.235 | 0.968 | 5.052 |
| regional | 615 | 0.307 | 0.233 | 1.058 | 1.616 |
| unknown (survey data) | 59 | 0.661 | 0.610 | 2.727 | 3.005 |
| *Incentive type* | | | | | |
| direct loan | 165 | 0.430 | 0.333 | 1.449 | 1.352 |
| loan guarantee | 56 | 0.482 | 0.393 | 1.394 | 1.746 |
| subsidy | 686 | 0.251 | 0.187 | 0.747 | 1.772 |
| tax credit | 47 | 0.468 | 0.426 | 3.590 | 13.354 |
| unspecified or mixed | 112 | 0.455 | 0.411 | 1.723 | 3.006 |
| *Methodology* | | | | | |
| Difference in differences | 150 | 0.253 | 0.207 | 0.934 | 2.664 |
| Regression discontinuity design | 137 | 0.328 | 0.263 | 0.746 | 2.053 |
| Matched difference in differences | 319 | 0.251 | 0.188 | 1.132 | 5.394 |
| Matching | 323 | 0.307 | 0.232 | 1.02 | 1.711 |
| Other | 137 | 0.591 | 0.504 | 1.888 | 1.571 |
| Directly targeted outcome | 292 | 0.425 | 0.336 | 1.855 | 5.647 |
| Indirectly targeted outcome | 774 | 0.283 | 0.224 | 0.839 | 1.893 |
| Simultaneous effect | 431 | 0.350 | 0.295 | 1.393 | 4.806 |
| Non simultaneous effect | 635 | 0.302 | 0.227 | 0.93 | 1.897 |
| *Estimate refers to* | | | | | |
| all firms (grand ATE or ATT) | 531 | 0.303 | 0.247 | 1.235 | 4.44 |
| disadvantaged firms | 183 | 0.464 | 0.377 | 1.291 | 2.166 |
| advantaged firms | 122 | 0.23 | 0.156 | 0.892 | 1.418 |
| other subgroup of firms | 230 | 0.3 | 0.226 | 0.827 | 1.732 |

## 4. The network of co-authorship

So far, we worked under the standard assumption that the study-level random coefficients are independent from one another. However, if one looks at the list of authors of the studies involved in our analysis (see Web Supplementary Material), it immediately emerges that the Italian literature tends to gather around a limited number of relatively prolific names. These authors may evaluate the same programme in multiple studies, although at different points in time or emphasising different aspects. In total, the studies under investigation can be ascribed to 74 authors, with 22 of them signing more than one study. In particular, 10 authors sign two, 7 sign three and 4 sign four studies. One single author signs twelve studies. Sometimes these prolific authors work with each other, other times they work in connection with other authors that sign only that specific study. Other times, there are isolated studies written by one-shot (co-)author(s).

Under these circumstances, the assumption of independence between study-level random coefficients, invoked in section 3.3, requires to be carefully assessed. To do so, we must envision some plausible departure from it. The most straightforward departure is that articles sharing at least one author may not be independent. Dependence might be due to an author's mindset, competencies and so forth that contribute to multiple studies, as well as to the fact that a same author may be using the same data multiple times. Under these circumstances, it seems sensible to consider directed linkages from earlier studies toward later ones, and bi-directed links between studies that were developed over the same time period (i.e., they appeared in the same calendar year or with a maximum lag of one year). We can visualise the resulting situation in Figure 2, using social network analysis tools (Wasserman and Faust, 1994; Scott and Carrington, 2011).

*Figure 2. Co-authorship network of the studies under investigation, where earlier studies influence later ones, and concomitant studies influence each other*

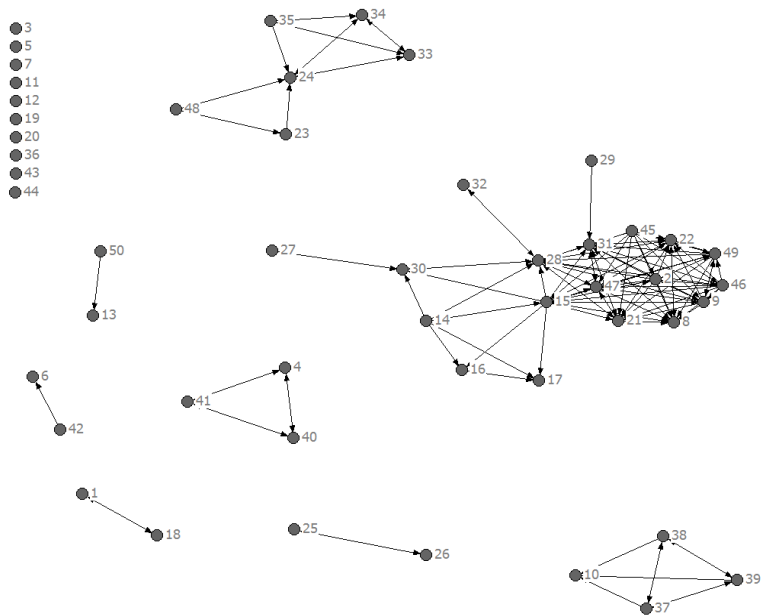

Notes. Nodes, marked with black circles, are articles, identified by the numerical identifier reported in the reference list. Lines are the co-authorship linkages that connect the nodes. Relationships are directed, or bi-directed. Network visualisation is performed using Ucinet (Borgatti et al., 2002).

The network structure can be described through a $J \times J$ asymmetric adjacency matrix $W$, whose elements $w_{hk}$ (h,k=1,…,J) are equal to 1 if study $h$ receives influence from study $k$, and 0 otherwise.

We can extend model [2] to allow for correlated random coefficients as follows. Let v be the results of a Simultaneously Autoregressive (SAR) process (Anselin, 1988): $v = \rho W v + u$, with unknown autocorrelation coefficient $\rho$ that quantifies the strength of the between-study dependence described in W (in row-standardised form) and, again, $u \sim N(0, \sigma_u^2 I_J)$. Then v can be expressed as $v = (I_J - \rho W)^{-1} u$ and model [2] becomes

$$\eta = X\beta + D(I_J - \rho W)^{-1} u. \qquad [3]$$

The random coefficients now follow a Normal distribution $v \sim N(0, \Sigma)$ where the covariance matrix $\Sigma$ is defined by two unknown parameters, $\rho$ and $\sigma_u^2$, that need to be estimated:

$$\Sigma = \sigma_u^2 \left[ (I_J - \rho W)'(I_J - \rho W) \right]^{-1}. \qquad [4]$$

From [4] it is easy to note that, if $\rho = 0$, the assumption of independence between study-level random coefficients holds. In such case, the appropriate model is the one introduced in Section III. Instead, if $\rho \neq 0$, the model accounting for between-study dependence is preferable.

## 5. Results

### 5.1 Appropriateness of our multilevel meta-regression model

Before discussing the main results of the analysis, we wish to highlight how the data support our choice to account for unobserved study heterogeneity through a multilevel model. Moreover, we wish to show that, in spite of the co-authorship network examined in Section 4, the usage of a model where study-level random coefficients are assumed to be independent from one another is statistically reasonable with the data at hand.

In order to establish whether unobserved study heterogeneity actually represents a non-negligible issue, we estimate the variance parameter $\sigma_u^2$ related to model [1]. Then we compare the deviance of the multilevel model to the deviance of an ordinary logit model that has the same covariates and test the difference against a Chi-bar distribution (see Snijders and Bosker, 2012, pp. 98-99). As shown in Table 4 (columns A and B), the test supports the appropriateness of a multilevel model. The Table also reports estimates of $\sigma_u$, i.e. the standard deviation of the random coefficients, which is anything but negligible in the models for both $\Pr(t_{ij} > 1.645)$ (column A) and $\Pr(t_{ij} > 1.96)$ (column B). In fact, the intraclass correlation, i.e. the proportion of total variance accounted for by the study-level random coefficients, is 25.8% in the former, and 26.9% in the latter.

In addition, we follow up the reasoning outlined in Section 4 and assess whether the assumption of independent study-level random coefficients is plausible in our context. Table 4, columns C and D, reports estimates of $\sigma_u$, $\rho$ and $\beta$ from a model where the vector of random terms is $v = \rho W v + u$. The estimate of $\rho$, which is expected to quantify the strength of the between-study dependence, is not statistically different from zero, whereas the estimate of the standard deviation of the residual independent random components, $\sigma_u$, is significantly positive. Furthermore, the estimates of $\beta$ are quite similar across columns A and C, as well as across columns B and D. Therefore, the hypothesis of independent study-level random coefficients seems plausible.

## 5.2 Probability of a positive effect for different types of programmes

Let us now comment on the estimated coefficients $\beta$ and, in parallel, use such coefficients, provided they are statistically significant, to predict probability differences across alternative levels of the explanatory variables. While predicting these values, we neutralise the influence of unobserved study heterogeneity by fixing each random effect $u_j$ at its mean value of zero. This allows to generalise the inference to all Italian programmes analogous to those analysed here, and to go beyond study-specific factors of success (or of failure). For the same reason, we set $\sqrt{n}=0$ and the remaining covariates at their mean value.

The coefficient is positive if the treatment effect estimate refers to an outcome that is directly targeted by the programme rather than to another outcome (Table 4). For instance, an R&D programme is more likely to succeed in raising private R&D expenditure than productivity or sales. This coefficient translates into a 30.1% higher probability of having a significant (at 5%) positive estimate (Table 5). With 95% confidence, such higher probability ranges between 18.1% and 42.1%, which leaves almost no doubts about the fact that the type of outcome chosen in the study makes a difference.

Moreover, we find a negative coefficient when the treatment effect estimate refers to an outcome that is lagged forward in time, rather than measured immediately after programme participation, which translates into an interval prediction of the differential probability of having a significant (at 5%) positive estimate between -34.4% and -6.4% (point prediction is -20.4%). Therefore, timing also matters. These results suggest the idea that enterprise and innovation programmes can be more effective in supporting the initial stages of the investment process than in ensuring that such process is completed with success, or in leading to other positive results later on.

The estimated coefficients also suggest that programmes are better at supporting disadvantaged firms, rather than advantaged ones. In fact, whereas both the positive coefficient we yield for disadvantaged firms, and the negative one we estimate for advantaged firms, are at times barely significant relative to the baseline category (all firms), the direct contrasts between disadvantaged and advantaged firms is characterised by an extremely significant coefficient (*p*-value = 0.001) equal to 1.11 in favour of the former. In fact, the differential probability of having a significant (at 5%) positive estimate is point predicted at 26%, with a confidence interval from 12.1% to 39.8% (Table 5). This result generalises to a broad set of programmes supporting business investment the finding achieved by Castellacci and Mee Lie (2015) in their MRA concerned with R&D tax credits alone. If one believes that policies should alleviate some of the constraints on investments faced by smaller and younger firms, rather than picking those who are already winners, then our finding indicates that, in Italy, these policies are far from being useless.

The coefficient associated with $\sqrt{n}$ is close to zero and statistically non-significant. This result suggests that the probability of having a significantly positive treatment effect estimate does not increase with larger study sample size, as would be expected if there was publication bias. To this regard, we may see that also the coefficient of the publication status is insignificant. Therefore, publication bias does not seem to pose serious threats in our study.

All the previous results are essentially confirmed by a random-intercept meta-regression model that has identical covariates, but where the response variable is 1 if the treatment effect estimate is significant at 2.5% (Table 4, Column B). Here, a significant positive coefficient is found if one shifts the programme goal from R&D to more indiscriminate investments, which translates into a 24.9% higher predicted probability of having a significant (at 2.5%) positive estimate (Table 5).

Unfortunately, the empirical evidence reported by the Italian literature at hand is not yet sufficient to draw conclusions on whether a significant positive effect is more or less likely to be met with national or regional programmes, or on whether one policy instrument works better than another one. However, as shown in Table 6, the probability of finding a positive treatment effect estimate may be rather high for the most common types of support schemes. The Table reports the predicted probability of an immediate positive effect for all firms, on an outcome that is directly targeted by the treatment and with respect to the most common of such schemes. For example, an R&D subsidy is expected to produce a significantly (at 5%) positive effect 60.2% of times, with an interval prediction ranging from 35.8% to 84.6%. An investment loan, instead, is expected to produce a significantly (at 5%) positive effect 86.4% of times; at worst, the effect is positive 70.6% of times.

*Table 4. Estimated model coefficients*

| | (A) | | (B) | | (C) | | (D) | |
|---|---|---|---|---|---|---|---|---|
| Outcome: | logit(Pr($t_{ij}$>1.645)) | | logit(Pr($t_{ij}$>1.96)) | | logit(Pr($t_{ij}$>1.645)) | | logit(Pr($t_{ij}$>1.96)) | |
| Model: | **Random intercept** | | **Random intercept** | | **Random intercept** | | **Random intercept** | |
| Study-level random coefficients $u_{ij}$ are: | assumed independent from each other | | | | assumed not independent from each other | | | |
| | Coefficient | S.E. | Coefficient | S.E. | Coefficient | S.E. | Coefficient | S.E. |
| *FIXED PART* | | | | | | | | |
| Programme aimed at investments (base: Aimed at R&D) | 0.736 | 0.570 | 1.232** | 0.615 | 0.747 | 0.754 | 1.198* | 0.871 |
| Directly targeted outcome (base: Indirectly targ. outc.) | 1.249*** | 0.265 | 0.896*** | 0.299 | 1.230*** | 0.29 | 0.884*** | 0.315 |
| Non simultaneous effect (base: Simultaneous) | -0.850*** | 0.301 | -0.688** | 0.343 | -0.853*** | 0.305 | -0.702* | 0.332 |
| Disadvantaged firms (base: All firms) | 0.446* | 0.259 | 0.446 | 0.285 | 0.444* | 0.282 | 0.453* | 0.302 |
| Advantaged firms (base: All firms) | -0.663** | 0.330 | -0.738* | 0.378 | -0.664* | 0.336 | -0.737* | 0.384 |
| Other subgroup of firms (base: All firms) | -0.489* | 0.284 | -0.595* | 0.315 | -0.486* | 0.312 | -0.589* | 0.343 |
| Loan guarantee (base: Direct loan) | -0.527 | 1.183 | -0.528 | 1.265 | -0.430 | 1.735 | -0.159 | 1.964 |
| Subsidy (base: Direct loan) | -0.697 | 0.532 | -0.286 | 0.654 | -0.864* | 0.587 | -0.400 | 0.747 |
| Tax credit (base: Direct loan) | 0.385 | 1.059 | 1.064 | 1.131 | 0.217 | 1.451 | 0.988 | 1.637 |
| Unspecified or mixed instrument (base: Direct loan) | 0.987 | 0.824 | 1.577* | 0.919 | 0.967 | 1.123 | 1.409 | 1.260 |
| Regional programme (base: National programme) | -0.062 | 0.583 | -0.049 | 0.600 | -0.603 | 0.787 | -0.695 | 0.875 |
| Unknown governance level (base: National programme) | 2.327*** | 0.857 | 2.475*** | 0.887 | 2.661* | 1.375 | 2.848* | 1.501 |
| $\sqrt{n}$ (centred) | 0.001 | 0.006 | 0.001 | 0.006 | -0.003 | 0.010 | -0.002 | 0.012 |
| RDD (base: DID) | 0.168 | 0.937 | 0.196 | 0.972 | 0.158 | 1.255 | 0.348 | 1.453 |
| Matched DID (base: DID) | -0.334 | 0.853 | -0.600 | 0.891 | -0.452 | 1.249 | -0.533 | 1.445 |
| Matching (base: DID) | -0.288 | 0.851 | -0.404 | 0.886 | -0.408 | 1.268 | -0.452 | 1.455 |
| Other methodology (base: DID) | -0.029 | 0.972 | -0.040 | 1.009 | -0.576 | 1.413 | -0.529 | 1.585 |

[cont]

[cont]

| | | | | | | | | |
|---|---|---|---|---|---|---|---|---|
| Study was published (base: Unpublished) | 0.019 | 0.674 | 0.168 | 0.709 | 0.042 | 0.863 | 0.110 | 0.978 |
| Study-level proportion of estimates on directly affected outcomes | -1.200 | 1.054 | -1.070 | 1.096 | -1.537 | 1.500 | -1.451 | 1.689 |
| Study-level proportion of estimates of non-simultaneous effects | 1.430** | 0.724 | 0.745 | 0.773 | 1.844* | 0.980 | 1.156 | 1.082 |
| Study-level proportion of estimates regarding disadvantaged firms | 1.327 | 1.361 | 0.944 | 1.355 | 0.316 | 1.909 | -0.204 | 2.098 |
| Study-level proportion of estimates regarding advantaged firms | -0.648 | 2.260 | -0.872 | 2.384 | -1.540 | 3.603 | -2.236 | 4.268 |
| Study-level proportion of estimates regarding other subgroups of firms | -1.098 | 1.072 | -1.625 | 1.131 | -1.251 | 1.405 | -1.975 | 1.577 |
| Grand intercept | -0.315 | 1.342 | -0.475 | 1.425 | 0.963 | 1.943 | 0.848 | 2.181 |
| **RANDOM PART** | | | | | | | | |
| $\sigma_u$ | 1.144*** | 0.245 | 1.211*** | 0.247 | 1.661*** | 0.421 | 1.886*** | 0.504 |
| $\rho$ | | | | | 0.081 | 0.377 | 0.253 | 0.370 |
| LR test vs. marginal model (based on a Chi-bar distribution) | 30.56*** | | 35.77*** | | - | | - | |
| Observations | 1,066 | | 1,066 | | 1,066 | | 1,066 | |
| AIC | 1,151.5 | | 1,006.1 | | - | | - | |
| Log Likelihood | -549.7 | | -477.1 | | - | | - | |

* $p<0.10$;  ** $p<0.05$;  *** $p<0.01$

*Table 5. Predicted probability differences for alternative levels of selected explanatory variables*

| | | | | Probability difference | 95% C.I. | |
|---|---|---|---|---|---|---|
| *Model for $Pr(t_{ij} >1.645)$* | | | | | | |
| Programme aims at: | Investment | vs. | R&D | 0.175 | -0.087 | 0.438 |
| Type of outcome: | Directly targeted outcome | vs. | Other outcome | 0.301 | 0.181 | 0.421 |
| Timing of effects: | Non simultaneous | vs. | Simultaneous effect | -0.204 | -0.344 | -0.064 |
| Estimate refers to: | Disadvantaged | vs. | Advantaged firm | 0.260 | 0.121 | 0.398 |
| | Advantaged | vs. | All firms | -0.149 | -0.287 | -0.011 |
| | Disadvantaged | vs. | All firms | 0.111 | -0.015 | 0.236 |
| | | | | | | |
| *Model for $Pr(t_{ij} >1.96)$* | | | | | | |
| Programme aims at: | Investment | vs. | R&D | 0.249 | 0.006 | 0.491 |
| Type of outcome: | Directly targeted outcome | vs. | Other outcome | 0.194 | 0.057 | 0.330 |
| Timing of effects: | Non simultaneous | vs. | Simultaneous effect | -0.142 | -0.286 | 0.002 |
| Estimate refers to: | Disadvantaged | vs. | Advantaged firm | 0.235 | 0.097 | 0.373 |
| | Advantaged | vs. | All firms | -0.133 | -0.258 | -0.008 |
| | Disadvantaged | vs. | All firms | 0.102 | -0.028 | 0.233 |

*Table 6. Predicted probability of immediate positive effect for all firms on an outcome that is likely to be directly affected by treatment with respect to six common programmes*

|  | Model for Pr($t_{ij}$ >1.645) | | | Model for Pr($t_{ij}$ >1.96) | | |
|---|---|---|---|---|---|---|
|  | **Probability** | **95% C.I.** | | **Probability** | **95% C.I.** | |
| R&D subsidy | 0.602 | 0.358 | 0.846 | 0.351 | 0.096 | 0.605 |
| R&D loan | 0.752 | 0.506 | 0.998 | 0.418 | 0.046 | 0.790 |
| R&D tax credit | 0.817 | 0.536 | 1.000 | 0.676 | 0.249 | 1.000 |
| Investment subsidy | 0.759 | 0.573 | 0.945 | 0.649 | 0.399 | 0.900 |
| Investment loan | 0.864 | 0.706 | 1.000 | 0.711 | 0.404 | 1.000 |
| Investment loan guarantee | 0.789 | 0.438 | 1.000 | 0.593 | 0.062 | 1.000 |

Although it is impossible to say with sufficient certainty which programme works best, we must acknowledge that these figures are high enough to stir some optimism on these programmes. They basically rule out the idea that all these different programmes are a complete waste of money and that everything would be as good without them, which is in line with the conclusions already reached by Dimos and Pugh (2016) with respect to R&D subsidies alone.

# 6. Concluding remarks

In this paper, we perform a multilevel meta-regression analysis of programme evaluations of enterprise and innovation policies that were implemented in Italy. We find that a positive effect of such policies is more likely to emerge when treatment effects are estimated on outcome variables that are measured immediately after programme participation and on outcomes that are directly targeted by the policies themselves. Indeed, depending on the type of programme, the probability of occurrence of positive treatment effects is higher when the outcome variables refer, for example, to R&D expenditures, amount of capital investment, receipt of favourable bank loans or lower interest rates, than when it refers to other indicators of firm performance, such as patenting activity, turnover, growth of productivity, profitability or, more in general, employees. Although positive effects on the latter type of outcomes are often highly desired by policymakers, they are unfortunately less likely to arise, perhaps as they require that a certain causal chain of events takes place after the treatment, a causal chain whose completion the treatment itself may be unable to guarantee. Evidently, these policies are likely to achieve in the short run some results for which they were designed, but they are also unlikely to bring about more complex ones, or to promote change over a longer time horizon.

Another important result that we find is that weaker firms that suffer from tighter investment constraints are most likely to benefit from positive effects, whereas support to stronger firms is more likely to translate into a non-significant impact.

At any rate, the main conclusion of this meta-analysis is that the cliché that most enterprise and innovation programmes are a complete waste of money has to be rejected. In fact, although the available data do not allow yet to establish which type of programme works best, our findings show that the probability of obtaining some positive effect is quite high for all types of schemes. Though this result may stir some optimism, some caution is required, as a positive treatment effect estimate is only a necessary, but not sufficient, condition for these programmes being ultimately value for money. Moreover, optimism is justified to the extent that these policies are not required to respond to purposes for which they were not designed. Indeed, while these policies can support various forms of firms' investment, they may fail to achieve more complex development goals.

# Supplementary Material

## List of evaluation studies included in the meta-regression analysis